\documentclass[12pt]{iopart}
\usepackage{graphicx}
\usepackage{amssymb}
\usepackage{psfrag}

\begin{document}
\title{Thermodynamics of rotating black branes in
Einstein-Born-Infeld-dilaton gravity}
\author{M. H. Dehghani$^{1,2}$,
S. H. Hendi$^{1}$, A. Sheykhi$^{1}$ and H. Rastegar Sedehi$^{1}$}
\address{$1$. Physics Department and Biruni Observatory, Shiraz University, Shiraz 71454, Iran}
\address{$2$. Research Institute for Astrophysics and Astronomy of Maragha (RIAAM), Maragha, Iran}
\eads{mhd\_at\_shirazu.ac.ir, hendi\_at\_mail.yu.ac.ir,
asheykhi\_at\_mail.uk.ac.ir, hamid\_rrs83\_at\_yahoo.com}

\begin{abstract}

In this paper, we construct a new class of charged, rotating solutions of $%
(n+1)$-dimensional Einstein-Born-Infeld-dilaton gravity with Liouville-type
potentials and investigate their properties. These solutions are neither
asymptotically flat nor (anti)-de Sitter. We find that these solutions can
represent black brane, with inner and outer event horizons, an extreme black
brane or a naked singularity provided the parameters of the solutions are
chosen suitably. We also compute temperature, entropy, charge, electric
potential, mass and angular momentum of the black brane solutions, and show
that these quantities satisfy the first law of thermodynamics. We find that
the conserved quantities are independent of the Born-Infeld parameter $\beta
$, while they depend on the dilaton coupling constant $\alpha$. We also find
the total mass of the black brane with infinite boundary as a function of
the entropy, the angular momenta and the charge and perform a stability
analysis by computing the heat capacity in the canonical ensemble. We find
that the system is thermally stable for $\alpha \leq 1$ independent of the
values of the charge and Born-Infeld parameters, while for $\alpha> 1$ the
system has an unstable phase. In the latter case, the solutions are stable
provided $\alpha \leq \alpha_{\max}$ and $\beta \geq \beta_{\min}$, where $%
\alpha_{\max}$ and $\beta_{\min}$ depend on the charge and the
dimensionality of the spacetime. That is the solutions are unstable for
highly nonlinear electromagnetic field or when the dilaton coupling constant
is large.

\end{abstract}
 \maketitle

\section{Introduction}

While on the gravity side, the Lovelock gravity coupled nonminimally to a
scalar dilaton field appears in the low energy limit of string theory \cite
{Wit1}, the Born-Infeld action coupled to a dilaton field, appears in the
low energy limit of open superstring theory on the electrodynamic side \cite
{Frad}. Although one can consistently truncate such models, the presence of
the dilaton field cannot be ignored if one consider coupling of the gravity
to other gauge fields, and therefore one remains with Einstein-Born-Infeld
gravity in the presence of a dilaton field. Some efforts have been done to
construct exact solutions of Einstein-dilaton gravity in the presence of
linear electromagnetic field. For example, exact static dilaton black hole
solutions of Einstein-Maxwell-dilaton (EMd) gravity have been considered in
Refs. \cite{CDB1,CDB2,MW,PW,CHM,Cai}. Exact rotating black holes solutions
with curved horizons have been obtained only for some limited values of the
coupling constant \cite{Fr,kun}. Recently, in the absence of dilaton
potential, a class of asymptotically flat rotating solution in $D$%
-dimensional EMd gravity have been constructed by embedding the $D$%
-dimensional Myers-Perry \cite{MP} solutions in $(D+1)$ dimensions, and
performing a boost with a subsequent Kaluza Klein reduction \cite{kunz}.
This procedure leads to EMd black holes in $D$ dimensions, for particular
values of the dilaton coupling constant. For general dilaton coupling, the
properties of rotating charged dilaton black holes only with infinitesimally
small charge \cite{Cas} or small angular momentum have been investigated in
four \cite{Hor2} and five \cite{SR1} dimensions . When the horizons are
flat, exact rotating solutions of EMd gravity with Liouville-type potential
have been constructed in four \cite{Deh1,Deh2} and $(n+1)$ dimensions \cite
{SDRP}.

In this paper, we want to consider the black brane solutions of
Einstein-dilaton gravity in the presence of nonlinear electromagnetic field.
The idea of the nonlinear electrodynamics was first introduced in 1934 by
Born and Infeld in order to obtain a finite value for the self-energy of
point-like charges \cite{BI}. Although it become less popular with the
introduction of QED, in recent years, the Born-Infeld action has been
occurring repeatedly with the development of superstring theory, where the
dynamics of D-branes is governed by the Born-Infeld action \cite{Frad,Cal}.
It has been shown that, charged black holes solutions in
Einstein-Born-Infeld gravity are less singular in comparing with the
Reissner-Nordstr\"{o}m solution. In other word, there is no
Reissner-Nordstr\"{o}m-type divergence term $Q^{2}/r^{2}$ in the metric near
the singularity while it exist only a Schwarzschild-type term $m/r$ \cite
{Car,Ban}. In the absence of dilaton field, exact solutions of
Einstein-Born-Infeld black holes with/without cosmological constant have
been constructed \cite{Gar,Tamaki,Dey,Cai2}, while the rotating solutions of
EBI gravity with flat horizons have been obtained in \cite{Deh3}. In the
presence of dilaton, exact solutions of the Einstein Born-Infeld dilaton
(EBId) gravity are known only in three dimensions \cite{YI}. Numerical
studies of the four-dimensional spherically symmetric solutions of EBId
gravity have been done in \cite{Tam}. Some factorized black hole solutions
in four-dimensional EBId theory have been discovered in \cite{SRM}. Till
now, exact rotating solution of EBId gravity for an arbitrary value of
coupling constant have not been constructed. In this paper, we first
introduce the action of $(n+1)$-dimensional EBId gravity in the presence of
a potential and obtain its field equations. Then, we construct the $(n+1)$%
-dimensional rotating black brane solution of this theory with cylindrical
or toroidal horizons in the presence of Liouville-type potential and
investigate the effect of dilaton and Born-Infeld fields on the stability of
the solutions.

The outline of our paper is as follows. In Sec. \ref{Field} we introduce the
action of the $(n+1)$-dimensional EBId gravity and obtain the field
equations. In Sec. \ref{Charged}, we construct the $(n+1)$-dimensional
rotating dilaton black branes with a complete set of rotation parameters in
the EBId theory and investigate their properties. In Sec. \ref{Therm}, we
obtain the conserved and thermodynamical quantities of the $(n+1)$%
-dimensional black brane solutions and show that these quantities satisfy
the first law of thermodynamics. We also perform a stability analysis and
show that the dilaton creates an unstable phase for the solutions. We finish
our paper with some closing remarks.

\section{General Formalism\label{Field}}

We consider the $(n+1)$-dimensional action in which gravity is coupled to
dilaton and Born Infeld fields with an action
\begin{eqnarray}
I_{G} &=&-\frac{1}{16\pi }\int_{\mathcal{M}}d^{n+1}x\sqrt{-g}\left( \mathcal{%
\ R}-\frac{4}{n-1}(\nabla \Phi )^{2}-V(\Phi )+L(F,\Phi )\right)  \nonumber \\
&&-\frac{1}{8\pi }\int_{\partial \mathcal{M}}d^{n}x\sqrt{-h}\Theta (h),
\label{Act}
\end{eqnarray}
where $\mathcal{R}$ is the Ricci scalar curvature, $\Phi $ is the dilaton
field, $V(\Phi )$ is a potential for $\Phi $ and $L(F,\Phi )$ is
\begin{equation}
L(F,\Phi )=4\beta ^{2}e^{4\alpha \Phi /(n-1)}\left( 1-\sqrt{1+\frac{%
e^{-8\alpha \Phi /(n-1)}F^{2}}{2\beta ^{2}}}\right)
\end{equation}
Here, $\alpha $ is a constant determining the strength of coupling of the
scalar and Born-Infeld fields, $F^{2}=F^{\mu \nu }F_{\mu \nu }$, where $%
F_{\mu \nu }=\partial _{\mu }A_{\nu }-\partial _{\nu }A_{\mu }$ is the
electromagnetic tensor field,$\ A_{\mu }$ is the vector potential, and $%
\beta $ is the Born-Infeld parameter with dimension of mass. In the limit $%
\beta \rightarrow \infty $, $L(F,\Phi )$ reduces to the standard Maxwell
field coupled to a dilaton field

\begin{equation}
L(F,\Phi )=-e^{-4\alpha \Phi /(n-1)}F^{2}.
\end{equation}
On the other hand, $L(F,\Phi )\rightarrow 0$ as $\beta \rightarrow 0$. It is
convenient to set
\begin{equation}
L(F,\Phi )=4\beta ^{2}e^{4\alpha \Phi /(n-1)}{\mathcal{L}}(Y),
\end{equation}
where
\begin{eqnarray}
&&{\mathcal{L}}(Y)=1-\sqrt{1+Y},  \label{LY} \\
&&Y=\frac{e^{-8\alpha \Phi /(n-1)}F^{2}}{2\beta ^{2}}.  \label{Y}
\end{eqnarray}
The last term in Eq. (\ref{Act}) is the Gibbons-Hawking boundary term which
is chosen such that the variational principle is well-defined. The manifold $%
\mathcal{M}$ has metric $g_{\mu \nu }$ and covariant derivative $\nabla
_{\mu }$. $\Theta $ is the trace of the extrinsic curvature $\Theta ^{ab}$
of any boundary(ies) $\partial \mathcal{M}$ of the manifold $\mathcal{M}$,
with induced metric(s) $h_{ab}$. In this paper, we consider the action (\ref
{Act}) with a Liouville type potential,
\begin{equation}
V(\Phi )=2\Lambda e^{4\alpha \Phi /(n-1)},  \label{v1}
\end{equation}
where $\Lambda $ is a constant which may be referred to as the cosmological
constant, since in the absence of the dilaton field ($\Phi =0$) the action (%
\ref{Act}) reduces to the action of Einstein-Born-Infeld gravity with
cosmological constant.

The equations of motion can be obtained by varying the action (\ref{Act})
with respect to the gravitational field $g_{\mu \nu }$, the dilaton field $%
\Phi $ and the gauge field $A_{\mu }$ which yields the following field
equations

\begin{eqnarray}
\mathcal{R}_{\mu \nu } &=&\frac{4}{n-1}\left( \partial _{\mu }\Phi \partial
_{\nu }\Phi +\frac{1}{4}g_{\mu \nu }V(\Phi )\right) -4e^{-4\alpha \Phi
/(n-1)}\partial _{Y}{\mathcal{L}}(Y)F_{\mu \eta }F_{\nu }^{\;\eta }
\nonumber \\
&&+\frac{4\beta ^{2}}{n-1}e^{4\alpha \Phi /(n-1)}\left[ 2Y\partial _{Y}{%
\mathcal{L}}(Y)-{\mathcal{L}}(Y)\right] g_{\mu \nu },  \label{FE1}
\end{eqnarray}
\begin{equation}
\nabla ^{2}\Phi =\frac{n-1}{8}\frac{\partial V}{\partial \Phi }+2\alpha
\beta ^{2}e^{4\alpha \Phi /(n-1)}\left[ 2Y \partial _{Y}{\mathcal{L}}(Y)-%
\mathcal{L}(Y)\right] ,  \label{FE2}
\end{equation}
\begin{equation}
\partial _{\mu }\left( \sqrt{-g}e^{-4\alpha \Phi /(n-1)}\partial _{Y}{%
\mathcal{L}}(Y)F^{\mu \nu }\right) =0.  \label{FE3}
\end{equation}
In particular, in the case of linear electrodynamics with $\mathcal{L}%
(Y)=-Y/2$, the system of equations (\ref{FE1})-(\ref{FE3}) reduce to the
well-known equations of EMd gravity \cite{CHM}.

\section{$(n+1)$-dimensional Solutions in EBId Gravity}

\label{Charged} \label{field}Our aim here is to construct the $(n+1)$%
-dimensional rotating solutions of the field equations (\ref{FE1})-(\ref{FE3}%
) with $k$ rotation parameters and investigate their properties. The
rotation group in $(n+1)$-dimensions is $SO(n)$ and therefore the number of
independent rotation parameters for a localized object is equal to the
number of Casimir operators, which is $[n/2]\equiv k$, where $[x]$ is the
integer part of $x$. The metric of $(n+1)$-dimensional rotating solution
with cylindrical or toroidal horizons and $k$ rotation parameters can be
written as \cite{awad,Lemos}

\begin{eqnarray}
ds^{2} &=&-f(r)\left( \Xi dt-{{\sum_{i=1}^{k}}}a_{i}d\phi _{i}\right) ^{2}+%
\frac{r^{2}}{l^{4}}R^{2}(r){{\sum_{i=1}^{k}}}\left( a_{i}dt-\Xi l^{2}d\phi
_{i}\right) ^{2}  \nonumber \\
&&-\frac{r^{2}}{l^{2}}R^{2}(r){\sum_{i<j}^{k}}(a_{i}d\phi _{j}-a_{j}d\phi
_{i})^{2}+\frac{dr^{2}}{f(r)}+\frac{r^{2}}{l^{2}}R^{2}(r)dX^{2},  \nonumber
\\
\Xi ^{2} &=&1+\sum_{i=1}^{k}\frac{a_{i}^{2}}{l^{2}},  \label{Met3}
\end{eqnarray}
where $a_{i}$'s are $k$ rotation parameters. The functions $f(r)$ and $R(r)$
should be determined and $l$ has the dimension of length which is related to
the cosmological constant $\Lambda $ for the case of Liouville-type
potential with constant $\Phi $. The angular coordinates are in the range $%
0\leq \phi _{i}\leq 2\pi $ and $dX^{2}$ is the Euclidean metric on the $%
(n-k-1)$-dimensional submanifold with volume $\Sigma _{n-k-1}$.

The modified Maxwell equation (\ref{FE3}) can be integrated immediately to
give
\begin{eqnarray}
F_{tr} &=&\frac{q\Xi e^{4\alpha \Phi /(n-1)}}{\left( rR\right) ^{n-1}\sqrt{1+%
{\frac{q^{2}}{\beta ^{2}\left( rR\right) ^{2n-2}}}}},  \nonumber  \label{Ftr}
\\
F_{\phi _{i}r} &=&-\frac{a_{i}}{\Xi }F_{tr},
\end{eqnarray}
where $q$, is an integration constant related to the electric charge of the
brane. In order to solve the system of equations (\ref{FE1}) and (\ref{FE2})
for three unknown functions $f(r)$, $R(r)$ and $\Phi (r)$, we make the
ansatz
\begin{equation}
R(r)=e^{2\alpha \Phi /(n-1)}.  \label{Rphi}
\end{equation}
Using (\ref{Rphi}), the Born-Infeld fields (\ref{Ftr}) and the metric (\ref
{Met3}), one can easily show that the field equations (\ref{FE1}) and (\ref
{FE2}) have solution of the form
\begin{eqnarray}
f(r) &=&\frac{2\left( \Lambda -2\beta ^{2}\right) (\alpha
^{2}+1)^{2}b^{2\gamma }}{(n-1)(\alpha ^{2}-n)}r^{2(1-\gamma )}-\frac{m}{%
r^{(n-1)(1-\gamma )-1}}  \nonumber \\
&&-\frac{4\beta ^{2}(\alpha ^{2}+1)b^{2\gamma }}{n-1}r^{(n-1)(\gamma
-1)+1}\int {r^{(n+1)(1-\gamma )-2}}\sqrt{1+\eta }{dr},  \label{f}
\end{eqnarray}
\begin{equation}
\Phi (r)=\frac{(n-1)\alpha }{2(1+\alpha ^{2})}\ln (\frac{b}{r}),  \label{phi}
\end{equation}
where
\begin{equation}
\eta =\frac{q^{2}b^{2\gamma (1-n)}}{\beta ^{2}r^{2(n-1)(1-\gamma )}}.
\label{Gamma}
\end{equation}
Here $b$ and $m$ are arbitrary constants and $\gamma =\alpha ^{2}/(\alpha
^{2}+1)$. The integral can be done in terms of hypergeometric function and
can be written in a compact form as
\begin{eqnarray}
f(r) &=&\frac{2\Lambda (\alpha ^{2}+1)^{2}b^{2\gamma }}{(n-1)(\alpha ^{2}-n)}%
r^{2(1-\gamma )}-\frac{m}{r^{(n-1)(1-\gamma )-1}}-\frac{4\beta ^{2}(\alpha
^{2}+1)^{2}b^{2\gamma }r^{2(1-\gamma )}}{(n-1)(\alpha ^{2}-n)}\times
\nonumber \\
&&\left( 1-\;_{2}F_{1}\left( \left[ -\frac{1}{2},\frac{\alpha ^{2}-n}{2n-2}%
\right] ,\left[ \frac{\alpha ^{2}+n-2}{2n-2}\right] ,-\eta \right) \right) .
\end{eqnarray}
One may note that as $\beta \longrightarrow \infty $ these solutions reduce
to the $(n+1)$-dimensional charged rotating dilaton black brane solutions
given in \cite{SDRP}. In the absence of a nontrivial dilaton ($\alpha
=\gamma =0$), the above solutions reduce to the asymptotically AdS charged
rotating black brane solutions of Einstein-Born-Infeld gravity presented in
\cite{Deh3}, and in the limit $\beta \rightarrow \infty $ and $\alpha =0$,
these solutions reduce to those of \cite{awad,Deh4}.

Using the fact that $_{2}F_{1}(a,b,c,z)$ has a convergent series expansion
for $|z|<1$, we can find the behavior of the metric for large $r$. This is
given by
\begin{eqnarray}
f(r) &=&\frac{2\Lambda (\alpha ^{2}+1)^{2}b^{2\gamma }}{(n-1)(\alpha ^{2}-n)}%
r^{2(1-\gamma )}+\frac{2(\alpha ^{2}+1)^{2}b^{-2(n-2)\gamma }q^{2}}{%
(n-1)(\alpha ^{2}+n-2)r^{2(n-2)(1-\gamma )}}  \nonumber \\
&&-\frac{m}{r^{(n-1)(1-\gamma )-1}}-\frac{(\alpha
^{2}+1)^{2}b^{-2(2n-3)\gamma }q^{4}}{2\beta ^{2}(n-1)^{2}(\alpha
^{2}+3n-4)r^{2(2n-3)(1-\gamma )}}.
\end{eqnarray}
The last term in the right hand side of the above expression is the leading
Born-Infeld correction to the AdS black brane with dilaton field.

\subsection*{Properties of the solutions}

In order to study the general structure of these solutions, we first look
for the curvature singularities in the presence of dilaton gravity. It is
easy to show that the Kretschmann scalar $R_{\mu \nu \lambda \kappa }R^{\mu
\nu \lambda \kappa }$ diverges at $r=0$, it is finite for $r\neq 0$ and goes
to zero as $r\rightarrow \infty $. Thus, there is an essential singularity
located at $r=0$. The spacetime is neither asymptotically flat nor (A)dS. As
in the case of rotating black hole solutions of the Einstein gravity, the
above metric given by (\ref{Met3}) and (\ref{f}) has both Killing and event
horizons. The Killing horizon is a null surface whose null generators are
tangent to a Killing field. It is easy to see that the Killing vector
\begin{equation}
\chi =\partial _{t}+{{{\sum_{i=1}^{k}}}}\Omega _{i}\partial _{\phi _{i}},
\label{Kil}
\end{equation}
is the null generator of the event horizon, where $\Omega _{i}$ is the $i$th
component of angular velocity of the outer horizon which may be obtained by
analytic continuation of the metric.

One can obtain the temperature and angular momentum of the event horizon by
analytic continuation of the metric. Setting $t\rightarrow i\tau $ and $%
a_{i}\rightarrow ia_{i}$ yields the Euclidean section of (\ref{Met3}), whose
regularity at $r=r_{+}$ requires that we should identify $\tau \sim \tau
+\beta _{+}$ and $\phi _{i}\sim \phi _{i}+\beta _{+}\Omega _{i}$, where $%
\beta _{+}$ and $\Omega _{i}$'s are the inverse Hawking temperature and the
angular velocities of the outer event horizon. One obtains
\begin{eqnarray}
T_{+} &=&\frac{f^{{\prime }}(r_{+})}{4\pi \Xi }=-\frac{(\alpha
^{2}+1)b^{2\gamma }r_{+}^{1-2\gamma }}{2\pi (n-1)\Xi }\left( \Lambda -2\beta
^{2}(1-\sqrt{1+\eta _{+}})\right)  \nonumber \\
&=&\frac{(n-\alpha ^{2})m}{4\pi \Xi (\alpha ^{2}+1)}{r}_{+}^{(n-1)(\gamma
-1)}-\frac{q^{2}(\alpha ^{2}+1)b^{2(2-n)\gamma }}{\pi \Xi (\alpha ^{2}+n-2)}%
r_{+}^{2(2-n)(1-\gamma )-1}\times  \nonumber \\
&&\;_{2}F_{1}\left( \left[ {\frac{1}{2},\frac{{n+\alpha }^{2}{-2}}{{2n-2}}}%
\right] ,\left[ {\frac{{3n+\alpha }^{2}{-4}}{{2n-2}}}\right] ,-\eta
_{+}\right) ,  \label{Temp}
\end{eqnarray}
\begin{equation}
\Omega _{i}=\frac{a_{i}}{\Xi l^{2}}.  \label{Om1}
\end{equation}
As one can see from Eq. (\ref{f}), the solution is ill-defined for $\alpha =%
\sqrt{n}$. The cases with $\alpha >\sqrt{n}$ and $\alpha <\sqrt{n}$ should
be considered separately. In the first case where $\alpha >\sqrt{n}$, as $r$
goes to infinity the dominant term is the second term, and therefore the
spacetime has a cosmological horizon for positive values of the mass
parameter, despite the sign of the cosmological constant $\Lambda $. In the
second case where $\alpha <\sqrt{n}$, as $r$ goes to infinity the dominant
term is the first term, and therefore there exist a cosmological horizon for
$\Lambda >0$, while there is no cosmological horizons if $\Lambda <0$ . It
is worth to mention that when there exists a cosmological horizon, the
solution does not present black hole. This is due to the fact that the
metric function $f(r)$ has only one real positive root. In the case of ($%
\alpha <\sqrt{n}$ and $\Lambda <0$) the spacetimes associated with the
solution (\ref{f}) exhibit a variety of possible casual structures depending
on the values of the metric parameters $\alpha $, $\beta $, $m$, $q$, and $%
\Lambda $. One can obtain the casual structure by finding the roots of $%
f(r)=0$. Unfortunately, because of the nature of the exponents in (\ref{f}),
it is not possible to find explicitly the location of horizons for an
arbitrary value of $\alpha $. But, we can obtain some information by
considering the temperature of the horizons.

Using the fact that the temperature of the extreme black brane is zero, it
is easy to show that the condition for having an extreme black hole is that
the mass parameter is equal to $m_{\mathrm{ext}}$, where $m_{\mathrm{ext}}$
is given as

\begin{eqnarray}
m_{\mathrm{ext}} &=&\frac{2(1+\alpha ^{2})^{2}b^{\gamma \left[ \gamma
-(n-2)(1-\gamma )\right] /(1-\gamma )}}{(\alpha ^{2}-n)(n-1)}\left( \frac{q_{%
\mathrm{ext}}^{2}}{\beta ^{2}\eta _{\mathrm{ext}}}\right) ^{[n(1-\gamma
)-\gamma ]/[2(n-1)(1-\gamma )]}\times  \nonumber \\
&&\left\{\Lambda -2\beta ^{2}\left[ 1-\;_{2}F_{1}\left( \left[ -{\frac{1}{2},%
\frac{{\alpha }^{2}{-n}}{{2n-2}}}\right] ,\left[ {\frac{{\alpha }^{2}+n{-2}}{%
{2n-2}}}\right] ,-\eta _{\mathrm{ext}}\right) \right] \right\},  \nonumber \\
\eta _{\mathrm{ext}} &=&\frac{\Lambda (\Lambda -4\beta ^{2})}{4\beta ^{4}}.
\label{mext}
\end{eqnarray}
Indeed the metric of Eqs. (\ref{Met3}) and (\ref{f}) has two inner and outer
horizons located at $r_{-}$ and $r_{+}$, provided the mass parameter $m$ is
greater than $m_{\mathrm{ext}}$, an extreme black hole in the case of $m=m_{%
\mathrm{ext}}$, and a naked singularity if $m<m_{\mathrm{ext}}$.

\section{The Conserved and Thermodynamics \label{Therm}Quantities of the
black branes}

In this section we calculate the action and the thermodynamic and conserved
quantities of the solutions. For asymptotically AdS solutions, the way that
one can calculate these quantities and obtain finite values for them is
through the use of the counterterm method inspired by AdS/CFT correspondence
\cite{Mal}. Although, in the presence of a nontrivial dilaton field, the
spacetime may not behave as either dS ($\Lambda >0$) or AdS ($\Lambda <0$),
one may use the counterterm method for computing the conserved quantities.
Since our solutions have flat boundary [$R_{abcd}(h)=0$], there exists only
one boundary counterterm
\begin{equation}
I_{ct}=-\frac{1}{8\pi }\int_{\delta \mathcal{M}}d^{n}x\sqrt{-h}\left[ \frac{%
(n-1)V(\Phi )}{(\alpha ^{2}-n)}\right] ^{1/2}.  \label{Ict}
\end{equation}
One may note that the counterterm has the same form as in the case of
asymptotically AdS solutions with zero curvature boundary when $\alpha $
goes to zero. Having the total finite action $I=I_G +I_{ct}$, one can use
the quasilocal definition \cite{BY} to construct a divergence free
stress-energy tensor. For the case of manifolds with zero curvature boundary
the finite stress-energy tensor is
\begin{equation}
T^{ab}=\frac{1}{8\pi }\left\{ \Theta ^{ab}-\Theta h^{ab}+\left[ \frac{%
(n-1)V(\Phi )}{(\alpha ^{2}-n)}\right] ^{1/2}h^{ab}\right\} .  \label{Stres}
\end{equation}
The first two terms in Eq. (\ref{Stres}) is the variation of the action (\ref
{Act}) with respect to $h_{ab}$, and the last term is the variation of the
boundary counterterm with respect to $h_{ab}$. Thus a conserved charge
\begin{equation}
Q_{\mathcal{\xi }}=\int_{\mathcal{B}}d^{n-1}S^{a}\mathcal{\xi }^{b}T_{ab},
\label{charge}
\end{equation}
can be associated with a spacelike surface $\mathcal{B}$ (with normal $n^{a}$%
), provided the boundary geometry has an isometry generated by a Killing
vector $\mathcal{\xi }^{b}$ \cite{BY}. For boundaries with timelike ($\xi
=\partial /\partial t$) and rotational ($\varsigma =\partial /\partial
\varphi $) Killing vector fields, one obtains the mass and angular momentum
as the conserved quantities. Denoting the volume of the boundary at constant
$t$ and $r$ by $V_{n-1}=(2\pi )^{k}\Sigma _{n-k-1}$, the mass and angular
momentum per unit volume $V_{n-1}$ of the black branes ($\alpha <\sqrt{n}$)
can be calculated by use of Eq. (\ref{charge}). We find
\begin{equation}
{M}=\frac{b^{(n-1)\gamma }}{16\pi l^{n-2}}\left( \frac{(n-\alpha ^{2})\Xi
^{2}+\alpha ^{2}-1}{1+\alpha ^{2}}\right) m,  \label{Mass}
\end{equation}
\begin{equation}
J_{i}=\frac{b^{(n-1)\gamma }}{16\pi l^{n-2}}\left( \frac{n-\alpha ^{2}}{%
1+\alpha ^{2}}\right) \Xi ma_{i}.  \label{Angmom}
\end{equation}
For $a_{i}=0$ ($\Xi =1$), the angular momentum per unit volume vanishes, and
therefore $a_{i}$'s are the rotational parameters of the spacetime. Notice
that the total mass and angular momentum per unit volume are similar to the
mass and angular momentum of $(n+1)$-dimensional EMd solutions presented in
\cite{SDRP}.

Next, we calculate the electric charge of the solutions. To determine the
electric field we should consider the projections of the electromagnetic
field tensor on special hypersurfaces. The normal to such hypersurfaces is
\begin{equation}
u^{0}=\frac{1}{N},\;\;\;u^{r}=0,\;\;\;u^{i}=-\frac{V^{i}}{N},
\end{equation}
where $N$ and $V^{i}$ are the lapse function and shift vector. Then the
electric field is $E^{\mu }=g^{\mu \rho }e^{\frac{-4\alpha \phi }{n-1}%
}F_{\rho \nu }u^{\nu }$, and the electric charge per unit volume $V_{n-1}$
can be found by calculating the flux of the electric field at infinity,
yielding
\begin{equation}
{Q}=\frac{\Xi q}{4\pi l^{n-2}}.  \label{Charge}
\end{equation}
The electric potential $U$, measured at infinity with respect to the
horizon, is defined by \cite{Gub}
\begin{equation}
U=A_{\mu }\chi ^{\mu }\left| _{r\rightarrow \infty }-A_{\mu }\chi ^{\mu
}\right| _{r=r_{+}},  \label{Pot1}
\end{equation}
where $\chi $ is the null generator of the horizon given by Eq. (\ref{Kil}).
One can easily show that the vector potential $A_{\mu }$ corresponding to
the electromagnetic tensor (\ref{Ftr}) can be written as
\begin{eqnarray}
A_{\mu } &=&\frac{qb^{(3-n)\gamma }}{\Upsilon r^{\Upsilon }}%
\;_{2}F_{1}\left( \left[ {\frac{1}{2},\frac{{n+\alpha }^{2}{-2}}{{2(n-1)}}}%
\right] ,\left[ {\frac{{3n+\alpha }^{2}{-4}}{{2(n-1)}}}\right] ,-\eta \right)
\nonumber  \label{vectorpot} \\
&&\times \left( \Xi \delta _{\mu }^{t}-a_{i}\delta _{\mu }^{i}\right)
\hspace{0.5cm}(\mathrm{no\;sum\;on}\;i),
\end{eqnarray}
where $\Upsilon =(n-3)(1-\gamma )+1$. Therefore, the electric potential may
be obtained as

\begin{equation}
U=\frac{qb^{(3-n)\gamma }}{\Xi \Upsilon {r_{+}^{\Upsilon }}}%
\;_{2}F_{1}\left( \left[ {\frac{1}{2},\frac{{n+\alpha }^{2}{-2}}{{2(n-1)}}}%
\right] ,\left[ {\frac{{3n+\alpha }^{2}{-4}}{{2(n-1)}}}\right] -\eta
_{+}\right) .  \label{Pot}
\end{equation}

Black hole entropy typically satisfies the so called area law of the entropy
\cite{Beck}. This near universal law applies to almost all kinds of black
holes and black branes in Einstein gravity \cite{hunt}. In our case, due to
the presence of dilaton field, we calculate the entropy through the use of
Gibbs-Duhem relation
\begin{equation}
S=\beta (\mathcal{M}-\Omega _{i}J_{i}-QU)-I  \label{Gibs}
\end{equation}
where $I$ is the finite total action evaluated on the classical solution.
Using Eqs. (\ref{Act}) and (\ref{Ict}), the finite total action per unit
volume $V_{n-1}$ can be calculated as
\[
I=\frac{\Xi b^{(n-1)\gamma }{r}_{+}^{(n-1)(1-\gamma )}}{4l^{n-2}}\left(
\frac{(n-1)\lambda -\frac{r_{+}^{n-\gamma n-4+3\gamma }}{(\alpha ^{2}-n+4)}%
\Upsilon }{(n-\alpha ^{2})\lambda -\frac{r_{+}^{n\gamma -3\gamma -n+2}}{%
(\alpha ^{2}+n-2)}\Upsilon }-1\right) ,
\]
where
\begin{eqnarray*}
\lambda &=&\frac{m}{4(1+\alpha ^{2})}, \\
\Upsilon &=&\frac{q^{2}(1+\alpha ^{2})}{b^{2\gamma (n-2)}}\;_{2}F_{1}\left( %
\left[ {\frac{1}{2},\frac{{n+\alpha }^{2}{-2}}{{2n-2}}}\right] ,\left[ {%
\frac{{3n+\alpha }^{2}{-4}}{{2n-2}}}\right] ,-\eta _{+}\right) .
\end{eqnarray*}
Now using Gibbs-Duhem relation and the thermodynamic and conserved
quantities, calculated before, one obtains
\begin{equation}
{S}=\frac{\Xi b^{(n-1)\gamma }r_{+}^{(n-1)(1-\gamma )}}{4l^{n-2}},
\label{Entropy}
\end{equation}
for the entropy per unit volume $V_{n-1}$. This shows that the entropy obeys
the area law for our case where the horizon curvature is zero.

\section{Thermodynamics of the black branes}

\label{stab}Now, we check the first law of thermodynamics for our solutions
in Einstein-Born-Infeld-dilaton gravity. In order to do this, we obtain the
mass $M$ as a function of extensive quantities $S$, $\mathbf{J}$ and $Q$.
Using the expression for the mass, the angular momenta, the charge, and the
entropy given in Eqs. (\ref{Mass}), (\ref{Angmom}), (\ref{Charge}), (\ref
{Entropy}) and the fact that $f(r_{+})=0$, one can obtain a Smarr-type
formula as
\begin{equation}
M(S,\mathbf{J},Q)=\frac{\left[ (n-\alpha ^{2})Z+\alpha ^{2}-1\right] J}{%
l(n-\alpha ^{2})\sqrt{Z(Z-1)}},  \label{Smar}
\end{equation}
where $J^{2}=\sum_{i}^{k}{J_{i}}^{2}$ and $Z=\Xi ^{2}$ is the positive real
root of the following equation
\begin{eqnarray}
&&\frac{n}{l^{2}}-\frac{4^{(\alpha ^{2}+n-2)/(n-1)}l^{[(n-2)\alpha
^{2}-2n+3]/(n-1)}\pi J}{(1+\alpha ^{2})b^{\alpha ^{2}}\sqrt{(Z-1)}Z^{(\alpha
^{2}-1)/(2n-2)}}S^{(\alpha ^{2}-n)/(n-1)}+  \nonumber \\
&&\frac{4\beta ^{2}}{n-1}\left[ 1-\;_{2}F_{1}\left( \left[ \frac{-1}{2},%
\frac{\alpha ^{2}-n}{2n-2}\right] ,\left[ \frac{\alpha ^{2}+n-2}{2n-2}\right]
,\frac{-\pi ^{2}Q^{2}}{\beta ^{2}S^{2}}\right) \right] =0 .
\end{eqnarray}
One may then regard the parameters $S$, $\mathbf{J}$, and $Q$ as a complete
set of extensive parameters for the mass $M(S,\mathbf{J},Q)$ and define the
intensive parameters conjugate to $S$, $\mathbf{J}$ and $Q$. These
quantities are the temperature, the angular velocities, and the electric
potential. It is a matter of forward calculations to obtain:
\begin{eqnarray}
T &=&\left( \frac{\partial M}{\partial S}\right) _{J,Q}=\left( \frac{%
\partial M}{\partial Z}\right) _{J,Q}\left( \frac{\partial Z}{\partial S}%
\right) _{J,Q}=-\frac{J}{Sl\sqrt{Z(Z-1)}}  \nonumber \\
&& \hspace{1cm}-\frac{\pi (1+\alpha ^{2})4^{(1-\alpha ^{2})/(n-1)}b^{\alpha
^{2}}Q^{2}Z^{(\alpha ^{2}-n)/(2n-2)}}{(\alpha ^{2}+n-2)l^{(n-2)(\alpha
^{2}-1)/(n-1)}S^{(\alpha ^{2}+2n-3)/(n-1)}}\mathcal{F},  \label{Tsmar}
\end{eqnarray}
\begin{equation}
\Omega _{i}=\left( \frac{\partial M}{\partial J_{i}}\right) _{S,Q}= \left(
\frac{\partial M}{\partial J_{i}}\right) _{S,Q}+\left( \frac{\partial M}{%
\partial Z}\right) _{S,Q}\left( \frac{\partial Z}{\partial J_{i}}\right)
_{S,Q}=\frac{J_{i}}{lJ}\sqrt{\frac{Z-1}{Z}},  \label{Jsmar}
\end{equation}
\begin{eqnarray}
U &=&\left( \frac{\partial M}{\partial Q}\right) _{J,S}=\left( \frac{%
\partial M}{\partial Z}\right) _{J,S}\left( \frac{\partial Z}{\partial Q}%
\right) _{J,S}=  \nonumber \\
&& \hspace{1cm} +\frac{\pi (1+\alpha ^{2})4^{(1-\alpha ^{2})/(n-1)}b^{\alpha
^{2}}QZ^{(\alpha ^{2}-n)/(2n-2)}}{(\alpha ^{2}+n-2)l^{(n-2)(\alpha
^{2}-1)/(n-1)}S^{(\alpha ^{2}+n-2)/(n-1)}} \mathcal{F},  \label{Usmar}
\end{eqnarray}
where $\mathcal{F}=\;_{2}F_{1}\left( \left[ {\frac{1}{2},\frac{\alpha ^{2}+{%
n-2}}{2n-{2}}}\right] ,\left[ {\frac{\alpha ^{2}+{3n-4}}{{2n-2}}}\right] ,-%
\frac{{\pi }^{2}Q^{2}}{\beta ^{2}S^{2}}\right)$. Using Eqs. (\ref{Angmom}), (%
\ref{Charge}) and (\ref{Entropy}), it is easy to show that the intensive
quantities calculated by Eqs. (\ref{Tsmar}), (\ref{Jsmar}) and (\ref{Usmar})
coincide with Eqs. (\ref{Temp}), (\ref{Om1}) and (\ref{Pot}). Thus, these
thermodynamics quantities satisfy the first law of thermodynamics
\begin{equation}
dM=TdS+{{{\sum_{i=1}^{k}}}}\Omega _{i}d{J}_{i}+Ud{Q} .
\end{equation}

Finally, we investigate the stability of charged rotating black brane
solutions of Born-Infeld-dilaton gravity in the canonical ensemble. In the
canonical ensemble, the charge and the angular momenta are fixed parameters,
and therefore the positivity of the heat capacity or $(\partial T/\partial
S)_{\mathbf{J},Q}=(\partial ^{2}M/\partial S^{2})_{\mathbf{J},Q}$ is
sufficient to ensure the local stability of the system. We, first, consider
the solutions for $\alpha \leq 1$. Numerical calculations show that the
black brane solutions are stable independent of the values of charge and
Born-Infeld parameters in any dimensions provided $\alpha \leq 1$ . To
achieve further understanding for $\alpha \leq 1$, we perform the stability
analysis for the uncharged case $q=0$. We obtain

\begin{figure}[ht]
\centering
{\ \includegraphics[width=7cm]{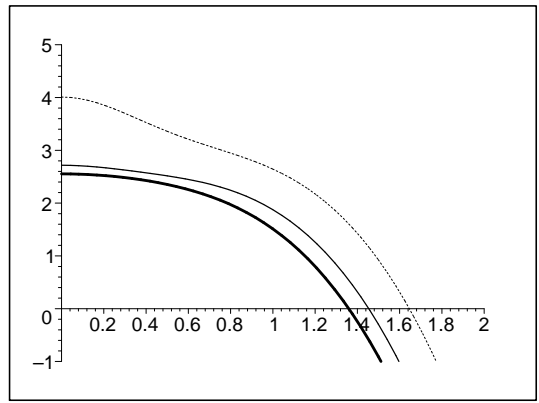} }
\caption{$(\partial ^{2}M/\partial S^{2})_{\mathbf{J},Q}$ versus $\protect%
\alpha $ for $\protect\beta =1$ and $n=5$. $q=0.1$ (bold line), $q=0.5$
(continues line), and $q=0.9$ (dashed line).}
\label{Figure1}
\end{figure}

\begin{figure}[ht]
\centering
{\ \includegraphics[width=7cm]{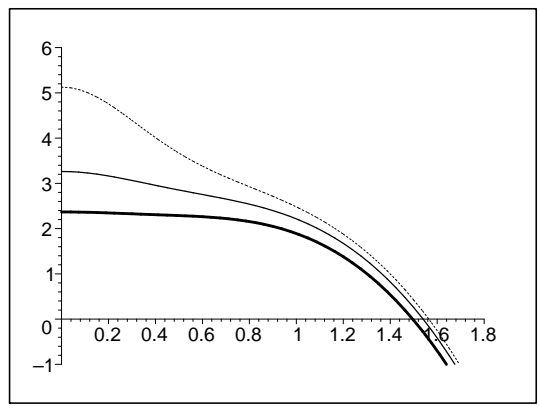} }
\caption{$(\partial ^{2}M/\partial S^{2})_{\mathbf{J},Q}$ versus $\protect%
\alpha $ for $n=5$ and $q=0.7$. $\protect\beta =0.5$ (bold line), $\protect%
\beta =1$ (continues line) and $\protect\beta =10$ (dashed line).}
\label{Figure2}
\end{figure}

\begin{figure}[ht]
\centering
{\ \includegraphics[width=7cm]{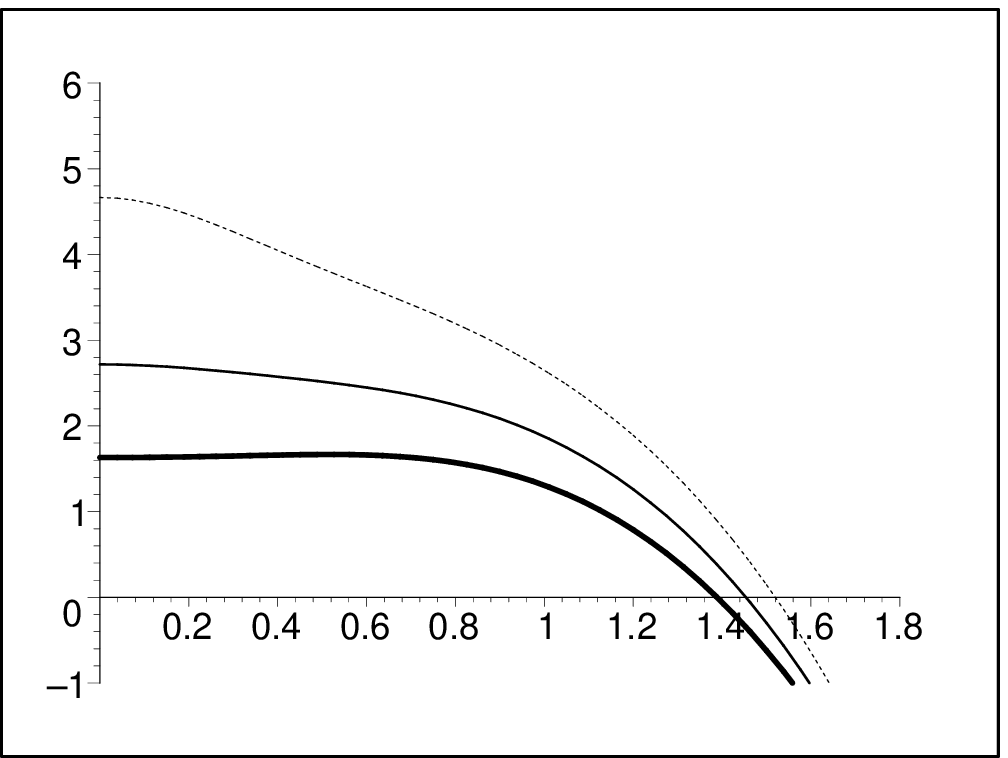} }
\caption{$(\partial ^{2}M/\partial S^{2})_{\mathbf{J},Q}$ versus $\protect%
\alpha $ for $q=0.5$ and $\protect\beta =1$. $n=4$ (bold line), $n=5$
(continues line), and $n=6$ (dashed line).}
\label{Figure3}
\end{figure}

\begin{figure}[ht]
\centering
{\ \includegraphics[width=7cm]{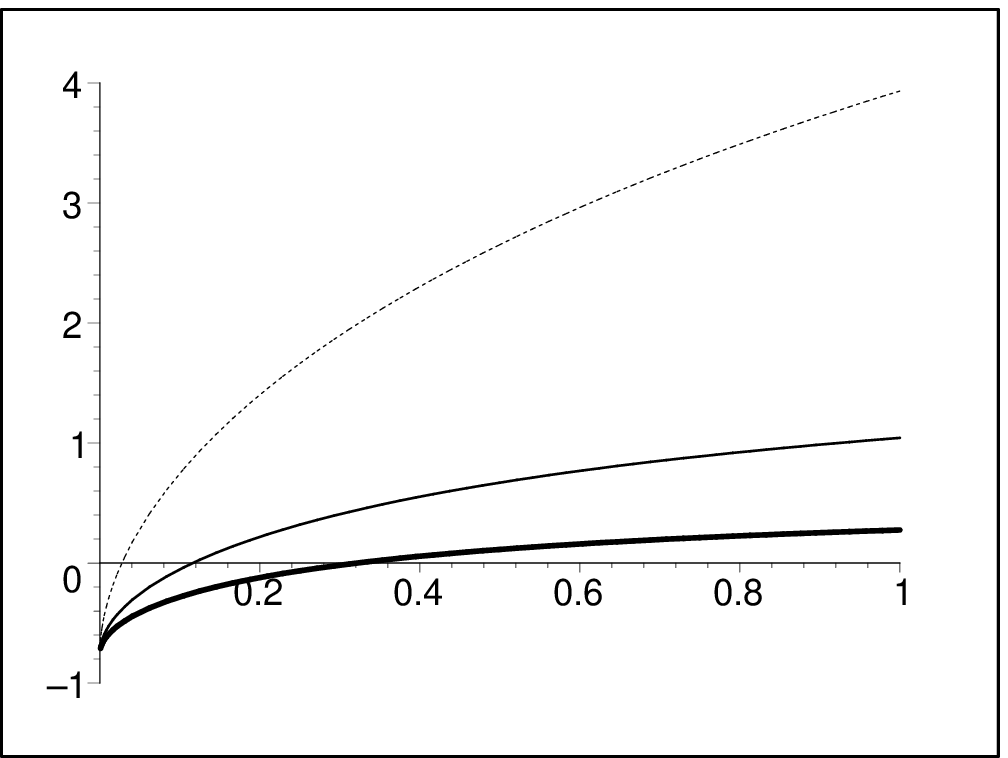} }
\caption{$(\partial ^{2}M/\partial S^{2})_{\mathbf{J},Q}$ versus $\protect%
\beta $ for $n=4$ and $\protect\alpha =\protect\sqrt{2}$. $q=0.7$ (bold
line), $q=1$ (continues line) and $q=2$ (dashed line).}
\label{Figure4}
\end{figure}

\begin{figure}[ht]
\centering
{\ \includegraphics[width=7cm]{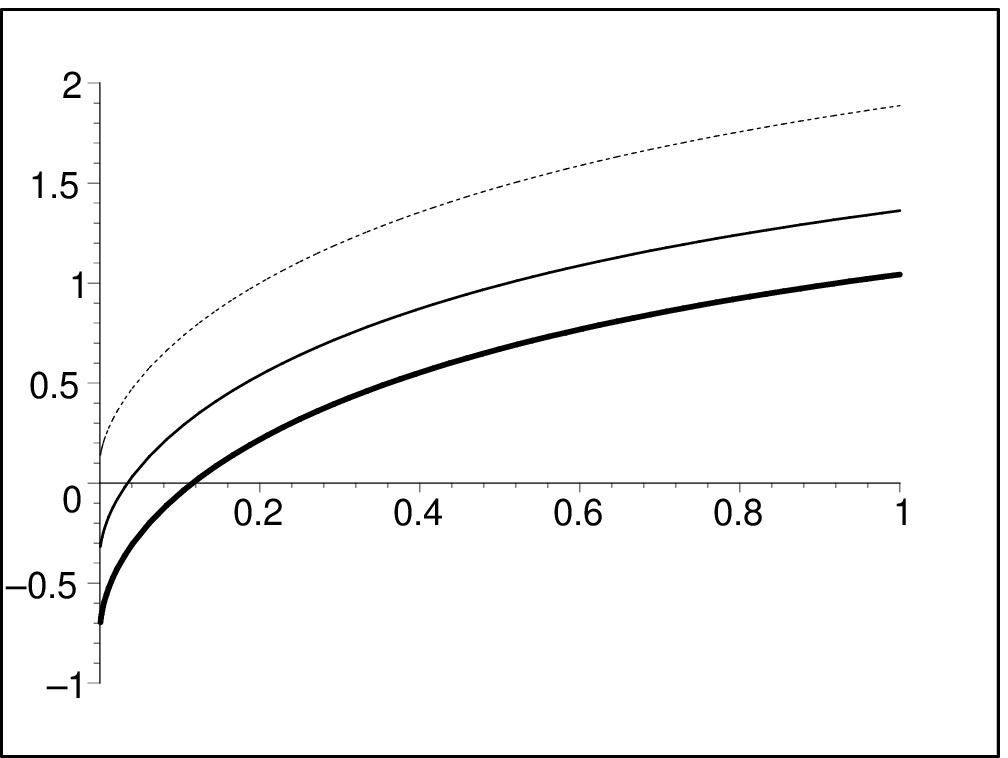} }
\caption{$(\partial ^{2}M/\partial S^{2})_{\mathbf{J},Q}$ versus $\protect%
\beta $ for $\protect\alpha =\protect\sqrt{2}$ and $q=1$. $n=4$ (bold line),
$n=5$ (continues line) and $n=6$ (dashed line).}
\label{Figure5}
\end{figure}

\begin{figure}[ht]
\centering
{\ \includegraphics[width=7cm]{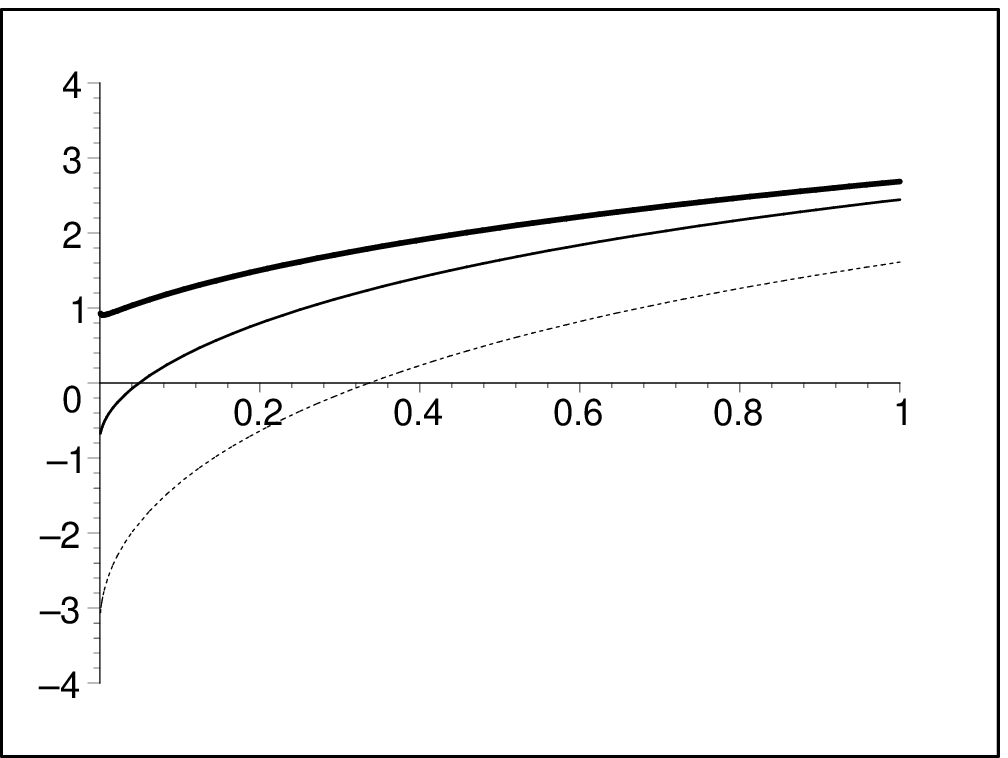} }
\caption{$(\partial ^{2}M/\partial S^{2})_{\mathbf{J},Q}$ versus $\protect%
\beta $ for $n=4$ and $q=1.5$. $\protect\alpha =1$ (bold line), $\protect%
\alpha =\protect\sqrt{2}$ (continues line) and $\protect\alpha =\protect%
\sqrt{3}$ (dashed line).}
\label{Figure6}
\end{figure}

\begin{figure}[ht]
\centering
{\ \includegraphics[width=7cm]{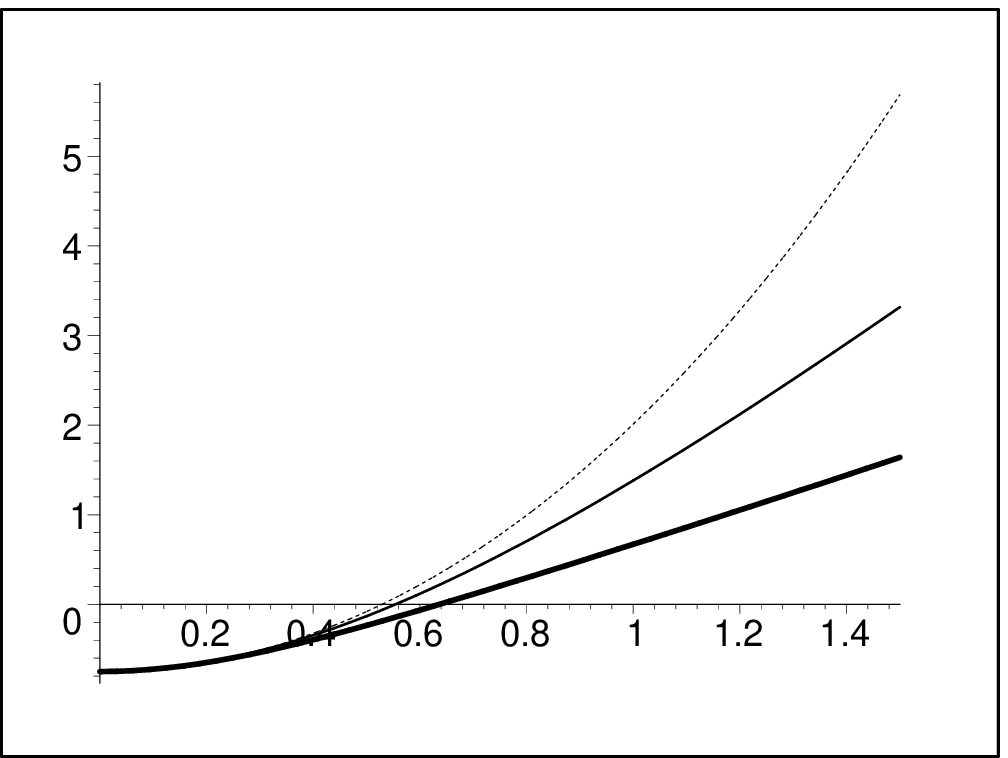} }
\caption{$(\partial ^{2}M/\partial S^{2})_{\mathbf{J},Q}$ versus $q$ for $%
\protect\alpha =\protect\sqrt{2}$ and $n=4$. $\protect\beta =0.5$ (bold
line), $\protect\beta =1$ (continues line) and $\protect\beta =10$ (dashed
line).}
\label{Figure7}
\end{figure}

\begin{figure}[ht]
\centering
{\ \includegraphics[width=7cm]{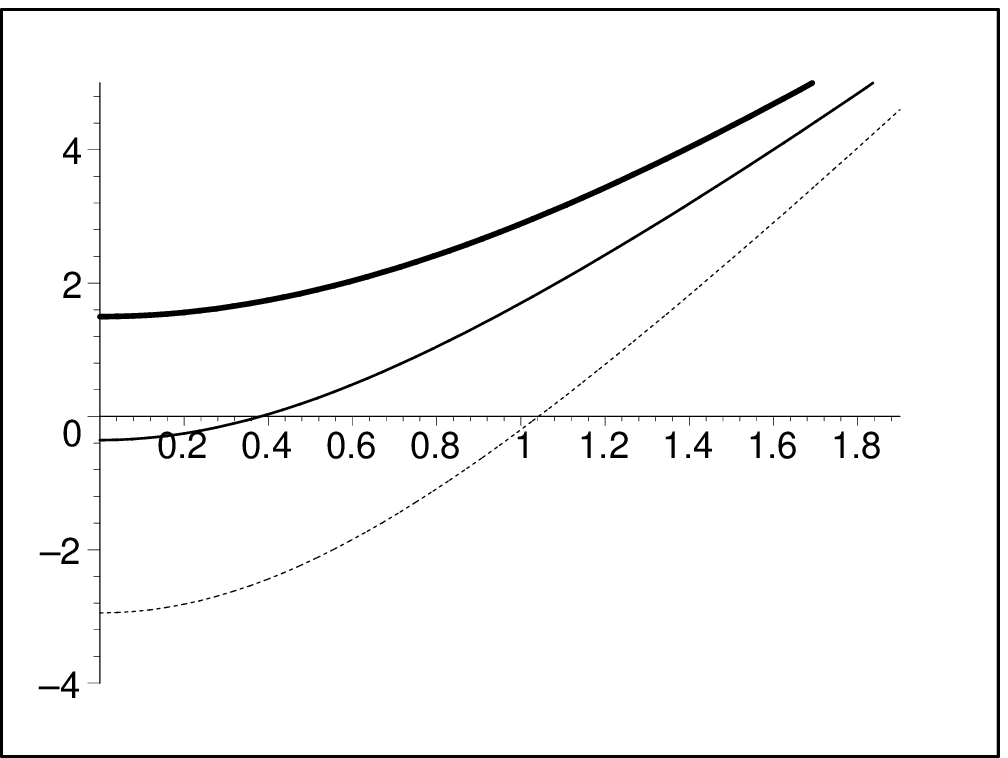} }
\caption{$(\partial ^{2}M/\partial S^{2})_{\mathbf{J},Q}$ versus $q$ for $%
n=5 $ and $\protect\beta=1$. $\protect\alpha=1$ (bold line), $\protect\alpha=%
\protect\sqrt{2}$ (continues line) and $\protect\alpha=\protect\sqrt{3}$
(dashed line).}
\label{Figure8}
\end{figure}

\begin{figure}[ht]
\centering
{\ \includegraphics[width=7cm]{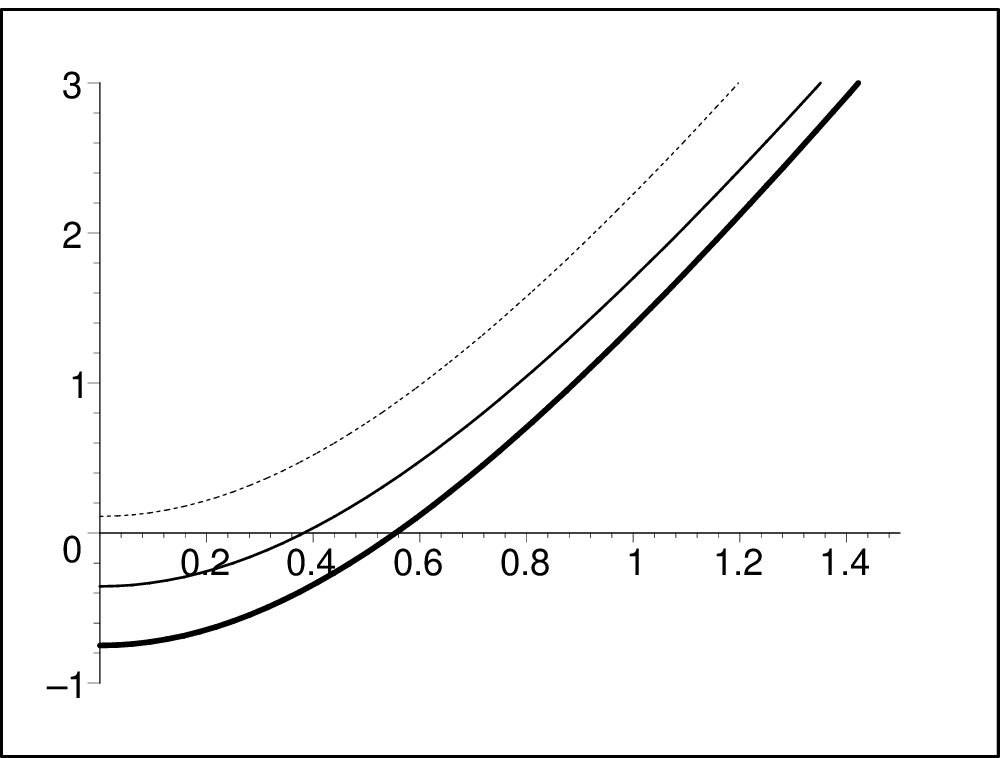} }
\caption{$(\partial ^{2}M/\partial S^{2})_{\mathbf{J},Q}$ versus $q$ for $%
\protect\alpha=\protect\sqrt{2}$ and $\protect\beta=1$. $n=4$ (bold line), $%
n=5$ (continues line) and $n=6$ (dashed line).}
\label{Figure9}
\end{figure}

\begin{eqnarray}
\left( \frac{\partial ^{2}M}{\partial S^{2}}\right) _{\mathbf{J}} &=&\frac{%
(\alpha ^{2}+1)l^{n-4}b^{-(n-3)\alpha ^{2}/(1+\alpha ^{2})}r_{+}^{-(n+\alpha
^{2}-2)/(1+\alpha ^{2})}}{\pi \Xi ^{2}\left[ \Xi ^{2}(n+\alpha
^{2}-2)+(1-\alpha ^{2})\right] }\times  \nonumber \\
&&\left[ (\Xi ^{2}-1)(n+1-2\alpha ^{2})+\Xi ^{2}(1-\alpha ^{2})\right] .
\label{dMSS}
\end{eqnarray}
Since $\Xi ^{2}\geq 1$, one finds that the above expression is positive
provided $\alpha \leq 1$. Thus, the uncharged solutions are stable for $%
\alpha \leq 1$. Also it is worth to mention that increasing the electric
charge not only does not change a stable solution to unstable one, but it
increases the stable phase of the system \cite{Deh4}. Thus, the charged
solutions are also stable for $\alpha \leq 1$. For $\alpha>1$, numerical
calculations show that the solutions may have an unstable phase. In order to
investigate the effects of the electric charge $q$, the dilaton coupling
constant $\alpha$, and the BI parameter $\beta$ on the unstable phase of the
solutions, we plot $(\partial ^{2}M/\partial S^{2})_{\mathbf{J},Q}$ versus $%
q $, $\alpha $ and $\beta $. Figures \ref{Figure1}, \ref{Figure2} and \ref
{Figure3} show that the solution is stable only for $\alpha \leq \alpha_{%
\mathrm{\max }}$, where $\alpha_{\mathrm{\max }}$ depends on $n$, $q$ and $%
\beta$. That is the dilaton field makes the solution unstable. Figure \ref
{Figure4} shows that as the nonlinearity of the electromagnetic field
increases ($\beta$ decreases), the stable phase of the solution decreases.
That is the nonlinearity of the electromagnetic field makes the solutions
more unstable. This fact may be seen in Figs. \ref{Figure4}, \ref{Figure5}
and \ref{Figure6} which show that the solutions are unstable for very highly
nonlinear field (small value of the Born-Infeld parameter). It is notable to
mention that $\beta _{\min }$ decreases with increasing $q$ or $n$ (see
Figs. \ref{Figure4} and \ref{Figure5}) and it increases with increasing $%
\alpha $ (see Fig. \ref{Figure6}). Also it is worth to mention that as in
the case of rotating solutions in the presence of linear electromagnetic
field, the solutions are unstable for small values of electric charge (see
Figs. \ref{Figure7}, \ref{Figure8} and \ref{Figure9}).

\section{Closing Remarks}

In this paper, we introduced the $(n+1)$-dimensional Einstein-Born-Infeld
action coupled to a dilaton field and obtained the equations of motion by
varying this action with respect to the gravitational field $g_{\mu \nu }$,
the dilaton field $\Phi $ and the gauge field $A_{\mu }$. In particular, in
the case of the linear electrodynamics with $\mathcal{L}(Y)=-{\frac{1}{2}}Y$%
, the system of equations reduce to the well-known equations of EMd gravity
\cite{CHM}. Then, we constructed a new class of charged rotating solutions
of $(n+1)$-dimensional Einstein-Born-Infeld-dilaton theory with cylindrical
or toroidal horizons in the presence of Liouville-type potentials and
investigated their properties. These solutions are neither asymptotically
flat nor (anti)-de Sitter. In the particular case $\beta \longrightarrow
\infty $, these solutions reduce to the $(n+1)$-dimensional charged rotating
dilaton black brane solutions constructed in Ref. \cite{SDRP}, while in the
absence of a nontrivial dilaton ($\alpha =\gamma =0$), the above solutions
reduce to the asymptotically AdS charged rotating black brane solutions of
Einstein-Born-Infeld theory presented in \cite{Deh3}. These solutions which
exist only for $\alpha^2 \neq n$ have only a cosmological horizon for (\emph{%
i}) $\alpha^2 >n$ despite the sign of $\Lambda $, and (\emph{ii}) positive
values of $\Lambda $, despite the magnitude of $\alpha $, and therefore they
do not represent black branes. For $\alpha^2 <n$, the solutions present
black branes with outer and inner horizons if $m>m_{ext}$, an extreme black
hole if $m=m_{ext}$, and a naked singularity if $m<m_{ext}$. We also
computed the action and thermodynamic quantities of the $(n+1)$-dimensional
rotating charged black brane such as the temperature, mass, angular
momentum, entropy, charge and electric potential, and found that they
satisfy the first law of thermodynamics. It is worth to mention that the
conserved quantities computed for when the boundary of spacetimes goes to
infinity.

Finally, we performed a stability analysis in the canonical ensemble by
considering $(\partial ^{2}M/\partial S^{2})_{\mathbf{J},Q}$ for the
rotating black brane solutions with infinite boundary in $(n+1)$-dimensional
EBId gravity. We found that there is no Hawking-Page phase transition in
spite of the presence of nonlinear electromagnetic and dilaton fields for $%
\alpha \leq 1$, while the solutions have an unstable phase for $\alpha>1$.
For $\alpha>1$, We found that there is always an upper limit $\alpha _{\max
} $ for which $(\partial ^{2}M/\partial S^{2})_{\mathbf{J},Q}$ is positive,
provided $\alpha <\alpha _{\max }$. It is worth to note that $\alpha _{\max}$
depend on the Born-Infeld parameter $\beta $, the charge $q$ and the
dimensionality of spacetime. On the other hand, we found a low limit for
Born-Infeld parameter $\beta _{\min }$, for which $(\partial ^{2}M/\partial
S^{2})_{\mathbf{J},Q}$ is positive provided $\beta >\beta _{\min }$. It is
notable to mention that $\beta _{\min }$ decreases with increasing $q$ or $n$%
, and it increases with increasing $\alpha $. Indeed, the solution is
unstable for highly nonlinear electromagnetic field. Note that the $(n+1)$%
-dimensional charged rotating solutions obtained here have flat horizons.
Thus, it would be interesting if one can construct charged rotating
solutions in $(n+1)$ dimensions in the presence of dilaton and Born-Infeld
fields with curved horizons.

\begin{ack}
This work has been supported by Research Institute for Astronomy
and Astrophysics of Maragha, Iran
\end{ack}


\begin{thebibliography}{99}
\bibitem{Wit1}  Green M B, Schwarz J H and Witten E, 1987 \textit{%
Superstring Theory}, (Cambridge: Cambridge University Press)

\bibitem{Frad}Fradkin E S and Tseytlin A A, 1985 \textit{Phys. Lett.} B
\textbf{163} 123 [SPIRES]  \\
Metsaev R R, Rakhmanov M A and Tseytlin A A, 1987
\textit{Phys. Lett.} B \textbf{193} 207 [SPIRES]\\
Bergshoeff E, Sezgin E, Pope C and Townsend P, 1987 \textit{Phys. Lett.} B
\textbf{188} 70 [SPIRES]

\bibitem{CDB1}  Gibbons G W and Maeda K, 1988 \textit{Nucl. Phys.} B \textbf{%
298} 741 [SPIRES]  \\
\noindent Koikawa T and Yoshimura M, 1987 \textit{Phys. Lett.} B
\textbf{189} 29 [SPIRES]\\
Brill D and Horowitz J, 1991 \textit{Phys. Lett.} B \textbf{262} 437 [SPIRES]

\bibitem{CDB2}  Garfinkle D, Horowitz G T and Strominger A, 1991 \textit{%
Phys. Rev.} D \textbf{43} 3140 [SPIRES]\\
Gregory R and Harvey J A, 1993 \textit{Phys. Rev.} D \textbf{47}
2411 [SPIRES] [hep-th/9209070]\\
Rakhmanov M A, 1994 \textit{Phys. Rev.} D \textbf{50} 5155
[SPIRES] [hep-th/9310174]

\bibitem{MW}  Mignemi S and Wiltshire D L, 1989 \textit{Class. Quantum Grav}%
. \textbf{6} 987 [SPIRES]\\
Wiltshire D L, 1991 \textit{Phys. Rev.} D \textbf{44} 1100
[SPIRES]\\
 Mignemi S and Wiltshire D L, 1992 \textit{Phys. Rev.} D
\textbf{46} 1475 [SPIRES] [hep-th/9202031]

\bibitem{PW}  Poletti S J and Wiltshire D L, 1994 \textit{Phys. Rev.} D
\textbf{50} 7260 [SPIRES] [gr-qc/9407021]

\bibitem{CHM}  Chan K C K, Horne J H and Mann R B, 1995 \textit{Nucl. Phys.}
B \textbf{447} 441 [SPIRES] [gr-qc/9502042]

\bibitem{Cai}  Cai R G, Ji J Y and Soh K S, 1998 \textit{Phys. Rev} D
\textbf{57} 6547 [SPIRES] [gr-qc/9708063]\\
Cai R G and Zhang Y Z, 2001 \textit{Phys. Rev.} D \textbf{64} 104015
[SPIRES] [hep-th/0105214]

\bibitem{Fr}  Frolov V P, Zelnikov A I and Bleyer U, 1987 \textit{Ann.
Phys., Lpz}. \textbf{44} 371\\
Belinsky V and Ruffini R, 1980 \textit{Phys. Lett.} B \textbf{89} 195
[SPIRES]

\bibitem{kun}  Kunduri H K and Lucietti J, 2005 \textit{Phys. Lett.} B
\textbf{609} 143 [SPIRES] [hep-th/0412153]\\
Yazadjiev S S, 2005 \textit{Phys. Rev.} D \textbf{72} 104014 [SPIRES]
[hep-th/0511016]

\bibitem{MP}  Myers R C and Perry M J, 1986 \textit{Ann. Phys.,} NY \textbf{%
172} 304 [SPIRES]

\bibitem{kunz}  Kunz J, Maison D, Navarro-Lerida F and Viebahn J, 2006
\textit{Phys. Lett.} B \textbf{639} 95 [SPIRES] [hep-th/0606005]

\bibitem{Cas}  Casadio R, Harms B, Leblanc Y and Cox P H, 1997 \textit{Phys.
Rev.} D \textbf{55} 814 [SPIRES] [hep-th/9606069]

\bibitem{Hor2}  Horne J H and Horowitz G T, 1992 \textit{Phys. Rev.} D
\textbf{46} 1340 [SPIRES] [hep-th/9203083]\\
Shiraishi K, 1992 \textit{Phys. Lett.} A \textbf{166} 298
[SPIRES]\\
Ghosh T and Mitra P, 2003 \textit{Class. Quantum Grav.}
\textbf{20} 1403 [SPIRES] [gr-qc/0212057]\\
Sheykhi A and Riazi N, 2006 \textit{Int. J. Theor. Phys.} \textbf{45} 2453
[SPIRES] [hep-th/0605072]

\bibitem{SR1}  Sheykhi A and Riazi N, \textit{Preprint} hep-th/0605042

\bibitem{Deh1}  Dehghani M H and Farhangkhah N, 2005 \textit{Phys. Rev.} D
\textbf{71} 044008 [SPIRES] [hep-th/0412049]

\bibitem{Deh2}  Dehghani M H, 2005 \textit{Phys. Rev.} D \textbf{71} 064010
[SPIRES] [hep-th/0411274]

\bibitem{SDRP}  Sheykhi A, Dehghani M H, Riazi N and Pakravan J, 2006
\textit{Phys. Rev.} D \textbf{74} 084016 [SPIRES] [hep-th/0606237]

\bibitem{BI}  Born M and Infeld L, 1934 \textit{Proc. R. Soc.} A \textbf{144}
425

\bibitem{Cal}  Callan C G, Lovelace C, Nappi C R and Yost S A, 1988 \textit{%
Nucl. Phys.} B \textbf{308} 221 [SPIRES]\\
Andreev O D and Tseytlin A A, 1988 \textit{Nucl. Phys.} B
\textbf{311} 205 [SPIRES]\\
Leigh R G, 1989 \textit{Mod. Phys. Lett.} A \textbf{4} 2767 [SPIRES]

\bibitem{Car}  Carlip S, 1995 \textit{J. Korean Phys. Soc.} \textbf{28} S447
[gr-qc/9503024]

\bibitem{Ban}  Banados M, Teitelboim C and Zanelli J, 1992 \textit{Phys.
Rev. Lett.} \textbf{69} 1849 [SPIRES] [hep-th/9204099]

\bibitem{Gar}  Garcia A, Salazar H and Plebanski J F, 1984 \textit{Nuovo.
Cim.} \textbf{84} 65 [SPIRES]\\
Demianski M, 1986 \textit{Found. Phys.} \textbf{16} 187\\
Breton N, 2003 \textit{Phys. Rev.} D \textbf{67} 124004 [SPIRES]
[hep-th/0301254]\\
Gibbons G W and Herdeiro C A R, 2001 \textit{Class. Quantum Grav.} \textbf{18%
} 1677 [SPIRES] [hep-th/0101229]

\bibitem{Tamaki}  Tamaki T, \textit{J. Cosmol. Astropart. Phys.}
JCAP05(2004)004 [SPIRES] [gr-qc/0310099]\\
Fernando S and Krug D, 2003 \textit{Gen. Rel. Grav.} \textbf{35}
129 [SPIRES] [hep-th/0306120]\\
Cataldo M and Garcia A, 1999 \textit{Phys. Lett.} B \textbf{456}
28 [SPIRES] [hep-th/9903257]\\
Fernando S, 2006 \textit{Phys. Rev.} D \textbf{74} 104032 [SPIRES]
[hep-th/0608040]

\bibitem{Dey}  Dey T K, 2004 \textit{Phys. Lett.} B \textbf{595} 484
[SPIRES] [hep-th/0406169]

\bibitem{Cai2}  Cai R G, Pang D W and Wang A, 2004 \textit{Phys. Rev.} D
\textbf{70} 124034 [SPIRES] [hep-th/0410158]

\bibitem{Deh3}  Dehghani M H and Rastegar Sedehi H, 2006 \textit{Phys. Rev.}
D \textbf{74} 124018 [SPIRES] [hep-th/0610239]

\bibitem{YI}  Yamazaki R and Ida D, 2001 \textit{Phys. Rev.} D \textbf{64}
024009 [SPIRES] [gr-qc/0105092]

\bibitem{Tam}  Tamaki T and Torii T, 2000 \textit{Phys. Rev.} D \textbf{62}
061501 [SPIRES] [gr-qc/0004071]\\
Tamaki T and Torii T, 2001 \textit{Phys. Rev.} D \textbf{64}
024027 [SPIRES] [gr-qc/0101083]\\
Clement G and Gal'tsov D, 2000 \textit{Phys. Rev.} D \textbf{62}
124013 [SPIRES] [hep-th/0007228]\\
Yazadjiev S S, Fiziev P P, Boyadjiev T L and Todorov M D, 2001 \textit{Mod.
Phys. Lett.} A \textbf{16} 2143 [SPIRES] [hep-th/0105165]

\bibitem{SRM}  Sheykhi A, Riazi N and Mahzoon M H, 2006 \textit{Phys. Rev.}
D \textbf{74} 044025 [SPIRES] [hep-th/0605043]\\
Yazadjiev S S, 2005 \textit{Phys. Rev.} D \textbf{72} 044006 [SPIRES]
[hep-th/0504152]

\bibitem{awad}  Awad A M, 2003 \textit{Class. Quantum Grav.} \textbf{20}
2827 [SPIRES] [hep-th/0209238]

\bibitem{Lemos}  Lemos J P S and Zanchin V T, 1996 \textit{Phys. Rev.} D
\textbf{54} 3840 [SPIRES] [hep-th/9511188]

\bibitem{Deh4}  Dehghani M H, 2002 \textit{Phys. Rev.} D \textbf{66} 044006
[SPIRES] [hep-th/0205129]\\
Dehghani M H and Khodam-Mohammadi A, 2003 \textit{Phys. Rev.} D \textbf{67}
084006 [SPIRES] [hep-th/0212126]

\bibitem{Mal}  Maldacena J, 1998 \textit{Adv. Theor. Math. Phys.} \textbf{2}
231 [SPIRES] [hep-th/9711200]\\
Witten E, 1998 \textit{Adv. Theor. Math. Phys.} \textbf{2} 253
[SPIRES] [hep-th/9802150]\\
Aharony O, Gubser S S, Maldacena J, Ooguri H and Oz Y, 2000
\textit{Phys. Rep.} \textbf{323} 183 [SPIRES] [hep-th/9905111]\\
Balasubramanian V and Kraus P, 1999 \textit{Commun. Math. Phys.} \textbf{208}
413 [SPIRES] [hep-th/9902121]

\bibitem{BY}  Brown J D and York J W, 1993 \textit{Phys. Rev.} D \textbf{47}
1407 [SPIRES] [gr-qc/9209012]

\bibitem{Gub}  Cvetic M and Gubser S S, 1999 \textit{J. High Energy Phys.}
JHEP04(1999)024 [SPIRES] [hep-th/9902195]\\
Caldarelli M M, Cognola G and Klemm D, 2000 \textit{Class. Quantum Grav.}
\textbf{17} 399 [SPIRES] [hep-th/9908022]

\bibitem{Beck}  Beckenstein J D, 1973 \textit{Phys. Rev.} D \textbf{7} 2333
[SPIRES]\\
Hawking S W, 1974 \textit{Nature} \textbf{248} 30 [SPIRES]\\
Gibbons G W and Hawking S W, 1977 \textit{Phys. Rev.} D \textbf{15} 2738
[SPIRES]

\bibitem{hunt}  Hunter C J, 1999 \textit{Phys. Rev.} D \textbf{59} 024009
[SPIRES] [gr-qc/9807010]\\
Hawking S W, Hunter C J and Page D N, 1999 \textit{Phys. Rev.} D
\textbf{59} 044033 [SPIRES] [hep-th/9809035]\\
Mann R B, 1999 \textit{Phys. Rev.} D \textbf{60} 104047 [SPIRES]
[hep-th/9903229]\\
Mann R B, 2000 \textit{Phys. Rev.} D \textbf{61} 084013 [SPIRES]
[hep-th/9904148]
\end{thebibliography}
\end{document}